\documentclass[traditabstract]{aa}  
\usepackage{graphicx}
\usepackage{natbib}
\usepackage{txfonts}
\newcommand{\Mv}{$\mathrm{M_{v}}$}

\newcommand{\Vt}{$\mathrm{V_{t}}$}
\newcommand{\Teff}{${T_{\rm eff}}$}
\newcommand{\logg}{$\log \mathrm{g}$}
\newcommand{\FeH}{$\mathrm{[Fe/H]}$}

\begin{document}

\authorrunning{Kovtyukh, Gorlova, Belik}
\titlerunning{Accurate luminosities for F--G supergiants}


 \title{Accurate luminosities from the oxygen $\lambda$7771--4 \AA\ triplet and
  the fundamental parameters of F--G supergiants}
 \author{V.~V.~Kovtyukh$^{1,2}$, N.~I.~Gorlova$^{3}$,
   \& S.~I.~Belik$^{1,2}$}
 \institute{Astronomical Observatory, Odessa National University, T.~G. Shevchenko Park, 65014, Odessa,
       Ukraine\\
    \email{val@deneb1.odessa.ua}
   \and
   Isaac Newton Institute of Chile, Odessa Branch, Ukraine
   \and
   Institute of Astronomy, Celestijnenlaan 200D, 3001, Leuven, Belgium
  }

 \abstract
  {The oxygen $\lambda$7771--4 \AA\ triplet is a good indicator of
  luminosity in A--G supergiants. However, its strength also depends on
  other atmospheric parameters.
  In this study, we present the luminosity calibrations
  where, for the first time, the effects of the effective temperature,
  microturbulent velocity, surface gravity,
  and the abundance have been disentangled.
  The calibrations are derived on the base of a dataset of high-dispersion
  spectra of 60 yellow supergiants with highly reliable luminosities and
  accurate atmospheric parameters.
  This allowed to bring the uncertainty of the triplet-based absolute
  magnitudes down to 0.26 mag.
  The calibrations are suitable for spectral types F0--K0 and luminosity
  classes I--II,
  covering absolute magnitudes \Mv\ from --1.0 to --10 mag.
 }
 {}


   \maketitle

\section{Introduction}

  F--G supergiants are among the most luminous stars in the Galaxy in
  the visual spectral region. Potentially, they can be used as
  standard candles for the extragalactic distance scale,
  for the studies of the spiral structure of our Galaxy
  and the evolution of massive stars (e.g. to establish
  the location of the blue loops on the Hertzsprung--Russell (HR) diagram).
  In practice, all these applications require knowledge of the accurate stellar luminosities.

  The oxygen O{\sc i} 7771-4 \AA\ triplet has been known as a luminosity
  indicator in supergiants for a long time (see \cite{me25, ke50}).
  Using the photoelectric technique on the sample of 10 F0--F8 supergiants with known luminosities
  (which in 8 cases were obtained from the distances to the parent clusters),
  \cite{os72} and \cite{ba74} derived the first quantitative relations
  between the strength of the triplet and the absolute magnitude (\Mv).
  Following that pioneering work, more extended studies have been
  carried out, both spectroscopically and photometrically
  \citep{so74, ka78, af89, af91, me93, af93, sl93, sl95}.
  Most recently, \cite{af03} presented calibrations based on the Hipparcos and Tycho data,
  that cover a wide range of \Mv\ with precision of 0.38 mag.
  \cite{dr09, dr12} and \cite{ne10, ne12} successfully used the oxygen triplet
  to distinguish F--G supergiants in M31, M33, the SMC and LMC, respectively,
  from the foreground Galactic dwarfs,
  with the goal to test the evolutionary tracks at a range of metallicities.

  The correlation of the triplet strength with luminosity is considered
  to be a result of the non-local-thermodynamical-equilibrium (non-LTE) effects,
  exacerbated by sphericity, that become
  stronger with higher luminosity  \citep{prz00}.
  However, as with any spectral feature, the strength of the triplet
  may also be a function of the effective temperature \Teff,
  microturbulent velocity \Vt, gravity \logg\, and metallicity [Fe/H].
  For example, the sensitivity to \Teff\ in the spectral range F5--G3
  was already noted by \cite{ke50} and \cite{fa88}.
  In the previous studies the number of supergiants with known luminosities
  was limited, which made impossible to take all these parameters into account.
  As a result, the precision of the early calibrations was rather low.
  The aim of the current work is to increase the accuracy of the luminosity
  determination based on a larger and better-characterized calibrating sample.

  In our first paper on the spectroscopic determination of the supergiant
  luminosities (\cite{ko10}, hereinafter Paper 1),
  we developed a method of the luminosity determination based on the
  ratios of the selected Fe{\sc i} and Fe{\sc ii} lines with different
  excitation potentials.
  We analyzed high-resolution spectra of 98 non-variable supergiants and
  53 classical Cepheids
  and derived 80 luminosity relations corresponding to different line pairs.
  Inclusion of Cepheids (in the specific pulsation phases) was a key factor
  to increase the number
  of objects with known luminosities, thanks to the period-luminosity
  relationship.
  The enlarged sample, in turn, resulted in a high precision of the
  line-ratio calibrations:
  $\sigma$(\Mv)=0.26 mag. For the present study, we selected only non-variable
  objects with the best quality data from that sample, and complemented
  them with the spectra of 12 new supergiants.

  In the era of the CCDs, the red spectral region containing the oxygen
  triplet is easily accessible, which is also assisted by the reddening
  of these population I stars.
  As a result, equally high S/N can be achieved for the oxygen 7771 triplet,
  as for the more traditional 5000--6800 \AA\, region used to measure the
  iron lines in Paper I.
  Then, the usage of the oxygen triplet as a luminosity calibrator may
  have several advantages over the iron lines.
  The triplet is usually stronger and hence can be measured in the
  lower $S/N$ spectra,
  as well as over the larger range of \Teff.
  While Fe{\sc ii} lines become unavailable at \Teff$<$4900 K,
  the triplet is measurable down to \Teff$=$4500 K.
  It is certainly faster to measure the three close lines than $\sim$60 iron lines
  that are spread over the whole visual spectral region.
  In the current study, we use the absolute magnitudes
  from the literature, including Paper I,
  to calibrate the strength of the oxygen triplet.
  In addition, while calibrations in Paper I considered only one other
  photospheric parameter --
  \Teff, in the current study, we also consider the dependence on
  the other three most important parameters --
  \Vt, \logg\, and [Fe/H], in attempt to increase the precision of the
  \Mv\ determination.

  We describe our observational material and the atmospheric parameters
  determination
  in Section 2. In Section 3, we present the calibration stars,
  derive new luminosity relations, and use them to
  evaluate \Mv\ for the entire sample of 74 supergiants.
  Section 4 summarizes our
  findings and provides suggestions for their application.

 \section{The Sample}

 \subsection{{\it The spectral material}}

  The spectra taken for the current study from Paper I
  mostly originate from the publicly available archive
  of the Ultraviolet-Visual Echelle-type Spectrograph (UVES), mounted
  on the Very Large
  Telescope (VLT) \citep{ba03}. All supergiants
  were observed in two instrumental modes, $Dichroic\,1$
  and $Dichroic\,2$, in order to provide almost complete coverage
  of the wavelength interval 3000-10\,000 \AA. The spectral resolution
  was about 80\,000, and the typical S/N ratio was 150--200 for the most of
  the spectra. The spectra were retrieved in the reduced form that was produced
  by the UVES pipe-line.

  Ten spectra were obtained using a coude-echelle  spectrometer \citep{mu99}
  mounted on the 2-m ``Zeiss'' telescope at the Peak Terskol Observatory
  located near Mt. Elbrus (Northern  Caucasus,  Russia).
  The resolution was set at R=52\,000.
  The observed wavelength range,
  $\lambda$3610--10\,270 \AA\ was covered by 86 echelle
  orders. The S/N reached 250 and more in the red part of the spectrum.
  The spectral extraction was performed by the authors using the dedicated
  software DECH95.

  We also included additional spectra for 27 supergiants.
  They were obtained with the fiber echelle-type
  spectrograph HERMES, mounted on the 1.2\,m Belgian
  telescope on La Palma. A high-resolution configuration with
  R= 85\,000 and the wavelength coverage 3800--9000 \AA\ was used.
  The spectra were reduced using the Python-based pipe-line
  that includes the order extraction, wavelength
  calibration with the Thr-Ne-Ar arcs, division by the flat
  field, cosmic-ray clipping, and the order merging.
  For more details on the spectrograph and the pipe-line, see
  \cite{ra11}.

  In all spectra, with a few exceptions, we measured two O{\sc i} indices:
  the equivalent width (EW) of the 7771 \AA\ line alone (EW71),
  which is the least blended member of the triplet,
  and the combined EW of all three components (EW74), see Fig. 1.
  It should be noted, that in the GK-type supergiants,
  the additional features of Fe{\sc i} at 7772.59 \AA\ and CN at 7772.9 \AA\
  become visible between the first (7771.95 \AA)
  and the second (7774.17 \AA) components of the triplet, contributing to
  the measured EW74 index. As will be shown in Sect. 3,
  it does not pose a problem for our calibrations.
  The EW71 index is blend-free in our spectra, which have resolution $>$50\,000.
  Whenever the red wing of the 7771.95 \AA\, component was affected, the blue wing
  was used for the Gaussian fitting and the EW determination.
  Therefore, the fact that the absolute magnitudes obtained
  from EW71 and EW74 turned out to agree very well,
  tells that the blends in the latter index are automatically calibrated together with the oxygen feature.
  Hence, our calibration of the EW74 index can be safely used in the low-resolution studies
  and the broad-line cases, at least above $T_{\rm{eff}}\approx 4500 K$.

  Our final sample of non-variable F--G supergiants
  with the O{\sc i} 7771-4 \AA\ triplet measured from the high-S/N, high resolution spectra,
  consists of 74 objects (Table 1), of which 60 have the luminosity estimates from the literature.
  To be able to use these stars as the spectroscopic luminosity
  calibrators, it is needed to obtain the accurate values of their
  atmospheric parameters.
  In the following sub-section, we describe how we derived them
  consistently for the whole sample.

  \begin{figure}
  \centering
  \includegraphics[width=9.0cm]{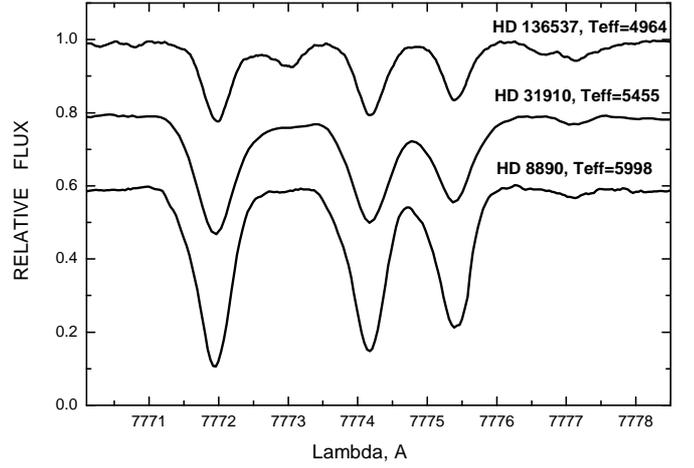}
  \caption{The 7771-4 triplet region in three supergiants with similar Mv $\approx$--3.0 but different $T_{\rm{eff}}$. For clarity, the bottom spectra have been shifted down along the flux axis.}
  \label{kovt1}
  \end{figure}
  %

  \begin{figure}
   \centering
   \includegraphics[width=9.0cm]{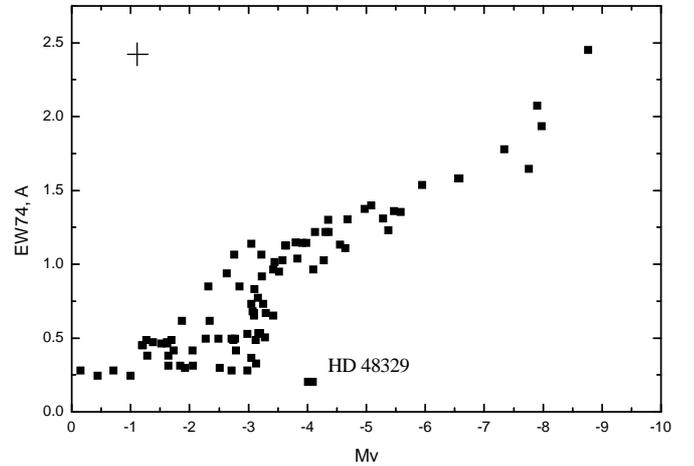}
  \caption{The equivalent width EW of the 7771-4 triplet vs. the absolute magnitude
  \Mv\ for our program stars from Table 1. Each data point represents an individual spectrum.
  The typical measurement error-bars are represented with a cross.}
  \label{kovt2}
  \end{figure}

 \subsection{{\it The atmospheric parameters}}

  The atmospheric parameters were determined
  following procedures outlined in the previous papers from this series
  \citep{ka99, ko08, ko10}.
  In short, the effective temperatures have been determined
  using the line-ratio calibrations from \cite{ko07}. The internal accuracy
  of the method is particularly high in the temperature range
  4500--7800 K where the standard deviation is of order 150 K or less,
  translating to 10--50 K in the standard error. One of the advantages
  of the line ratio method (or any other spectroscopic method) is its
  independence of the reddening, which is especially relevant for our
  supergiant stars that typically reside in the Galactic plane.

  The microturbulent velocities $V_{\rm t}$ and the surface gravities
  $\log g$ were derived using the modified approach,
  as described in \cite{ka99}. Namely, the microturbulence was determined
  from the requirement of no dependence on the EW of the iron abundance
  obtained from the individual Fe{\sc ii}, rather than Fe{\sc i}, lines.
  The surface gravity was established by forcing the iron
  abundance obtained from the Fe{\sc ii} lines, to be the same as
  the abundance derived from the {\it extrapolated to $EW=0$}
  (i.e., weak) Fe{\sc i} lines.
  The reason why strong Fe{\sc i} lines should be avoided
  is their sensitivity to the non-LTE effects \citep{lyu83, the99, mash11}.

  To obtain the elemental abundances, we used the WIDTH9 code by R. Kurucz
  and his atmospheric models, interpolated when needed \citep{ku92}.
  Those models have been calculated with one value of the microturbulent
  velocity of 4 km s$^{-1}$.
  To test what effect it can have on the abundances, we
  performed test calculations on the Cepheids, where the microturbulent
  velocity varies as a function of the pulsation phase. We found that
  the variation of several km s$^{-1}$ still has a negligible impact
  on the resulting abundances.
  We use the same oscillator strengths as in the previous papers from
  this series, in particular, by \cite{ka99}.
  They are based on the inverted analyses of the solar spectrum
  by adopting the solar abundances of \cite{gns96}.
  With this approach, the abundances derived for our program stars
  have typical uncertainty of $\pm0.15$ in [Fe/H] and $\pm0.20$ in [O/H].

The Oxygen abundances reported in Table 1 were derived using
the forbidden [O{\sc i}] 6300, 6363 \AA\ lines and the weak O{\sc i} 6156-8 \AA\ lines.
The former lines are free, while the latter are almost free, from the non-LTE effects,
unlike the $\lambda$7771-4 \AA\ triplet \citep{as04,fab09,prz00}.
One can still ask how the effects specific to the extended atmospheres
of supergiants, such as sphericity, may affect our abundance determination procedure in general.
On one hand, \cite{hau99} found a significant impact of sphericity
on line strengths in the yellow supergiants,
though the effect was considerably less than for the red supergiants,
whose spectra are dominated by the molecular bands.
On the other hand, \cite{prz00} based on the study of B-A supergiants, and \cite{Lobel00}
based on the M supergiant $\alpha$Ori, drew attention to the strong macro-turbulent
broadening that can smear the difference between the line profiles
calculated in the spherical and plane-parallel cases.
The issue clearly needs to be further investigated.
Until then, we simply rely on the weak lines for the abundance determination
 ($\leq$ 170 m\AA),
based on the calculations that on average, they form deeper and over the smaller extent
of the atmosphere, thus being the least susceptible to the non-LTE effects.
The existence of the empirical correlation of the strong
$\lambda$7771-4 \AA\ triplet with luminosity must be a manifestation of the fact,
that one or several of these poorly studied non-LTE effects
become systematically stronger at higher luminosities.

  The luminosities for the calibrating stars in this work
  are taken from Paper I, namely, the original
  values from the literature (based on the Hipparcos parallaxes and
  cluster distances)
  and the values we derived from the iron line ratios.
  Whenever more than one estimate was available, they were averaged
  with the equal weights.
  Table 1 lists our complete sample with the adopted atmospheric parameters,
  the oxygen EWs (in \AA), and the newly derived absolute magnitudes (see Sect. 3).
  Whenever more than one spectrum (up to five) was available for a given
  object, the atmospheric parameters and the widths of the triplet,
  inferred from the individual spectra,
  have been averaged. For objects with additional, more recent spectra,
  this resulted in the improved values, that can be slightly different
  from those in Paper I.
  Unlike Paper I, the current study concerns only non-variable supergiants.
  Cepheids were omitted because of the complicated behavior of
  $\log g$ and $V_{\rm t}$ within the pulsation cycle,
  and because of the possible influence of the dynamical processes on the non-LTE effects,
  which may affect the derived luminosities.

  \begin{figure}
  \includegraphics[width=9.0cm]{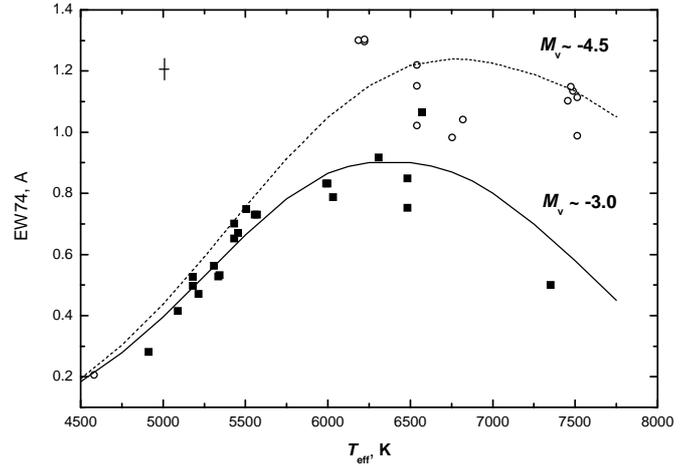}
  \caption{The dependence of the triplet's EW on the \Mv\ and \Teff.
     Squares represent objects with --3.5$<$\Mv$<$--2.5, circles with
     $-5.0<$\Mv$<-4.0$.
     Lines are drawn to show the approximate behavior with \Teff.
     More parameters are required to explain the remaining scatter.}
  \label{kovt3}
  \end{figure}
\begin{table*}
\caption{Calibrator supergiants, their parameters and calculated \Mv}
\label{T4}
\begin{tabular}{llrrrrrrlrrl}
\hline
\hline
  Star       & \Mv (literature)               &\Teff  & \logg& \Vt  &[Fe/H]&[O/H]    & EW71 & EW74  & \Mv (1)  & \Mv (2) & Comments \\
\hline
HD 000571 &  --3.05(1)                      &  6561 &  2.5&  3.3&--0.09 &   ...&   ... & 1.138 &   ...&--3.51&          \\
HD 000611 &  --2.50(3)                      &  5453 &  1.9&  3.2&  0.11 &   ...& 0.257 &   ... &--2.55&   ...& double   \\
HD 000725 &  --4.97(3)                      &  6793 &  1.8&  4.8&--0.16 &--0.23& 0.470 & 1.375 &--5.19&--5.28&          \\
HD 001457 &                                 &  7636 &  2.3&  4.8&--0.04 &--0.34& 0.482 & 1.350 &--5.59&--5.70&          \\
HD 003421 &  --1.62(1), --1.38(3)           &  5310 &  2.3&  2.6&--0.13 &--0.17& 0.148 & 0.472 &--1.45&--1.45&          \\
HD 004362 &  --3.07(3)                      &  5301 &  1.6&  4.4&--0.15 &   ...& 0.229 & 0.683 &--3.59&--3.63&          \\
HD 006130 &                                 &  7458 &  2.9&  5.0&  0.09 &--0.28& 0.399 & 1.102 &--4.97&--4.93&          \\
HD 007927 &  --8.76(2)                      &  7341 &  1.0&  8.7&--0.24 &--0.32& 0.930 & 2.453 &--9.16&--8.93&          \\
HD 008890 &  --3.10(3)                      &  5998 &  2.1&  4.7&  0.10 &  0.11& 0.316 & 0.832 &--2.93&--2.88& var      \\
HD 008906 &  --4.68(3)                      &  6710 &  2.2&  4.8&--0.07 &--0.42& 0.470 & 1.304 &--4.79&--4.71&          \\
HD 008992 &  --3.44(3)                      &  6234 &  2.4&  5.0&--0.02 &--0.23& 0.376 & 1.016 &--3.30&--3.24&          \\
HD 009167 &                                 &  7632 &  1.9&  6.0&--0.15 &  ... & 0.568 & 1.574 &--6.27&--6.41&          \\
HD 009900 &                                 &  4529 &  1.7&  2.7&  0.10 &--0.05& 0.043 & 0.274 &--2.51&--3.23&          \\
HD 009973 &  --5.08(3)                      &  6654 &  2.0&  5.7&--0.05 &--0.28& 0.541 & 1.399 &--5.35&--5.09&  Em      \\
HD 010494 &  --7.34(2)                      &  6672 &  1.2&  7.5&--0.20 &--0.32& 0.640 & 1.778 &--6.77&--6.76& NGC 654  \\
HD 011544 &  --3.28(3)                      &  5124 &  1.4&  3.5&  0.01 &  0.01& 0.169 & 0.504 &--3.46&--3.46&          \\
HD 012533 &                                 &  4319 &  1.5&  3.0&  0.04 &  ... & 0.070 & 0.146 &--3.58&--3.24&  SB      \\
HD 012545 &                                 &  4359 &  1.5&  2.0&--0.22 &  ... & 0.148 & 0.345 &--4.64&--3.82&  RSCVn   \\
HD 014662 &  --3.52(3)                      &  6111 &  2.2&  5.0&--0.10 &--0.21& 0.352 & 0.950 &--3.24&--3.23&  var     \\
HD 016088 &                                 &  7476 &  2.9&  3.3&  0.03 &--0.37& 0.408 & 1.147 &--4.94&--4.96&          \\
HD 016901 &  --3.25(2), --3.05(3)           &  5568 &  1.6&  4.2&  0.02 &--0.12& 0.262 & 0.731 &--3.27&--3.25& NGC 1039(?)  \\
HD 017971 &  --6.58(2), --6.56(3)           &  6822 &  1.3&  8.7&--0.20 &--0.41& 0.562 & 1.582 &--6.69&--6.82& IC 1848      \\
HD 018391 &  --7.76(2)                      &  5762 &  1.2& 11.5&--0.02 &--0.15& 0.638 & 1.647 &--7.71&--7.83& anon cluster \\
HD 020123 &  --2.06(1), --1.64(2), --1.84(3)&  5160 &  2.0&  3.3&--0.05 &  0.20& 0.130 & 0.314 &--2.25&--2.19& Melotte 20, SB\\
HD 020902 &  --4.13(1), --4.31(3), --4.36(4)&  6541 &  2.0&  4.8&--0.01 &--0.20& 0.430 & 1.218 &--4.38&--4.34& $\alpha$ Per cluster \\
HD 023230 &  --2.63(1)                      &  6560 &  2.4&  3.5&  0.04 &--0.09&   ... & 0.937 &   ...&--3.25& var,  vsini=44\\
HD 025291 &  --4.64(1)                      &  7400 &  2.6&  4.1&  0.00 &--0.31& 0.398 & 1.110 &--4.96&--4.97&          \\
HD 026630 &  --3.20(1), --3.17(3)           &  5309 &  1.8&  3.7&  0.02 &   ...& 0.200 & 0.533 &--2.85&--2.78& SB       \\
HD 031910 &  --3.29(1), --3.09(3)           &  5455 &  1.7&  4.5&--0.06 &--0.17& 0.233 & 0.671 &--3.28&--3.33& double   \\
HD 036673 &                                 &  7500 &  2.3&  4.4&  0.07 &--0.27& 0.435 & 1.210 &--5.29&--5.28&          \\
HD 036891 &  --2.79(3)                      &  5089 &  1.7&  3.3&--0.06 &--0.29& 0.130 & 0.415 &--2.77&--2.86&          \\
HD 045348 &  --5.37(1)                      &  7557 &  2.2&  2.7&--0.10 &--0.24& 0.442 & 1.231 &--5.29&--5.37& Canopus  \\
HD 048329 &  --4.09(3), --4.01(1)           &  4583 &  1.2&  3.7&  0.16 &   ...& 0.091 & 0.205 &--4.00&--3.82& var      \\
HD 054605 &  --7.97(2)                      &  6564 &  1.5& 10.2&--0.03 &--0.47& 0.725 & 1.936 &--7.43&--7.32& SB Coll121 var\\
HD 057118 &                                 &  7427 &  1.5&  7.5&--0.15 &--0.45& 0.671 & 1.823 &--7.22&--7.22&          \\
HD 065228 &  --1.87(1), --2.34(3)           &  5740 &  2.0&  3.9&  0.01 &--0.12& 0.226 & 0.616 &--2.43&--2.44&          \\
HD 067594 &  --3.12(1), --2.74(3)           &  5190 &  1.7&  3.7& -0.06 &--0.03& 0.172 & 0.488 &--3.04&--3.03&          \\
HD 070761 &                                 &  7460 &  1.9&  6.4&--0.04 &--0.31& 0.609 & 1.670 &--6.63&--6.62&          \\
HD 074395 &  --2.27(1), --2.77(3)           &  5247 &  1.8&  3.0&--0.01 &--0.07& 0.185 & 0.495 &--2.58&--2.45&          \\
HD 075276 &                                 &  6934 &  1.4&  3.9&--0.33 &--0.20& 0.508 & 1.388 &--5.71&--5.67&          \\
HD 077912 &  --2.51(1), --1.92(3)           &  4951 &  2.0&  2.4&  0.01 &--0.02& 0.090 & 0.299 &--1.92&--2.00& peculiar  \\
HD 079698 &  --0.15(1), --0.71(3)           &  5227 &  2.5&  1.7&  0.06 &--0.09& 0.106 & 0.281 &--0.67&--0.54&          \\
HD 084441 &  --1.28(1), --1.64(3)           &  5284 &  2.0&  2.2&  0.01 &--0.13& 0.127 & 0.382 &--1.54&--1.50& var      \\
HD 092125 &  --1.52(1), --1.62(3)           &  5336 &  2.4&  2.7&  0.05 &  0.10&   ... & 0.463 &   ...&--1.31&          \\
HD 101947 &  --7.90(2)                      &  6578 &  1.3& 11.3&--0.08 &--0.27& 0.811 & 2.075 &--8.26&--8.01& Stock 14, var\\
HD 109379 &  --0.44(1), --1.00(3)           &  5144 &  2.3&  1.7&  0.01 &--0.09& 0.079 & 0.245 &--0.87&--0.86&  var     \\
HD 117440 &                                 &  4741 &  1.9&  2.7&--0.01 &  0.04& 0.060 & 0.250 &--2.20&--2.50&  var     \\
HD 119605 &  --1.69(1), --1.27(3)           &  5441 &  2.1&  2.5&--0.28 &--0.07& 0.170 & 0.487 &--1.66&--1.61&          \\
HD 125809 &  --3.13(3)                      &  4845 &  2.1&  3.8&  0.00 &  0.02& 0.096 & 0.327 &--2.59&--2.80&          \\
HD 135153 &  --4.55(1)                      &  7488 &  3.1&  3.8&  0.01 &--0.22& 0.405 & 1.133 &--4.92&--4.94&          \\
HD 136537 &  --3.05(3)                      &  4964 &  1.3&  2.9&--0.06 &--0.24& 0.114 & 0.365 &--3.20&--3.29&          \\
HD 146143 &  --3.64(1), --3.62(3)           &  6068 &  2.0&  5.0&--0.13 &--0.29& 0.406 & 1.126 &--3.59&--3.55&          \\
HD 159181 &  --2.49(3), --2.71(1)           &  5180 &  2.1&  3.4&  0.03 &  0.03& 0.162 & 0.497 &--2.33&--2.34&          \\
HD 164136 &  --2.85(1), --2.32(3)           &  6483 &  3.1&  4.5&--0.37 &  0.05& 0.299 & 0.849 &--2.80&--2.93&          \\
HD 171635 &  --3.58(3), --4.28(1)           &  6151 &  2.1&  5.2&--0.04 &--0.15& 0.378 & 1.026 &--3.49&--3.46&          \\
HD 172594 &  --5.95(3)                      &  6849 &  1.5&  4.5&--0.15 &--0.29& 0.539 & 1.538 &--5.77&--5.77&          \\
HD 173638 &  --5.28(3)                      &  7444 &  2.4&  4.7&  0.11 &--0.31&   ... & 1.309 &   ...&--5.46&          \\
HD 174464 &  --3.83(3)                      &  6821 &  2.4&  4.8&--0.22 &--0.34& 0.377 & 1.040 &--4.40&--4.43&          \\
HD 180028 &  --3.23(3)                      &  6307 &  1.9&  4.0&  0.10 &--0.08& 0.340 & 0.917 &--3.44&--3.33&          \\
HD 182835 &  --5.58(3)                      &  6969 &  1.6&  4.9&  0.00 &--0.23& 0.421 & 1.353 &--5.21&--5.53&          \\
HD 185758 &  --1.21(3), --1.19(1)           &  5367 &  2.4&  2.1&--0.03 &--0.10&   ... & 0.452 &   ...&--0.96&          \\
HD 187203 &  --3.16(3)                      &  5710 &  2.2&  5.1&  0.05 &   ...&   ... & 0.772 &   ...&--2.86& post-AGB?\\
 \hline
\end{tabular}
\end{table*}

\begin{table*}
{Table 1. Continued}\\
\begin{tabular}{llrrrrrrlrrl}
\hline
\hline
  Star       & \Mv (literature)               &\Teff  & \logg& \Vt  &[Fe/H]&[O/H]    & EW71 & EW74  & \Mv (1)  & \Mv (2) & Comments \\
\hline
HD 193370   &  --3.80(3)                      &  6369 &  2.4&  6.1&  0.02 &--0.33&   ... & 1.149 &   ...&--3.99& SB       \\
HD 194093   &  --4.35(3)                      &  6188 &  1.7&  6.1&  0.05 &--0.09& 0.479 & 1.300 &--4.62&--4.54& var      \\
HD 195295   &  --2.76(1), --3.22(3)           &  6572 &  2.4&  3.5&  0.01 &--0.14& 0.386 & 1.065 &--3.65&--3.51& var      \\
HD 202240   &                                 &  7515 &  3.1&  3.7&  0.10 &  0.23& 0.336 & 1.112 &--4.46&--4.84&          \\
HD 204075   &  --2.05(1), --1.73(3)           &  5262 &  2.0&  2.6&--0.08 &--0.14& 0.128 & 0.416 &--1.81&--1.86& SB       \\
HD 204867   &  --3.09(1), --3.42(3)           &  5431 &  1.6&  4.1&--0.04 &--0.27& 0.227 & 0.652 &--3.25&--3.26&          \\
HD 206859   &  --2.98(1), --2.71(3)           &  4912 &  1.2&  2.5&  0.04 &   ...& 0.078 & 0.281 &--2.93&--3.11& var      \\
HD 209750   &  --2.98(1), --3.18(3)           &  5182 &  1.4&  3.5&  0.01 &--0.21& 0.180 & 0.527 &--3.42&--3.41&          \\
HD 231195   &                                 &  7032 &  1.0&  8.0&--0.09 &--0.19& 0.780 & 2.089 &--8.12&--7.89&          \\
HD 236433   &  --3.98(2), --3.91(3)           &  6541 &  2.2&  5.5&--0.15 &--0.46& 0.408 & 1.144 &--4.32&--4.35&NGC 129, SB\\
BD +37 3827 &  --5.47(3)                      &  6830 &  1.9&  6.0&--0.15 &--0.23& 0.488 & 1.360 &--5.49&--5.51&           \\
BD +60 2532 &  --4.10(2), --3.42(3)           &  6268 &  1.8&  5.2&--0.01 &--0.15& 0.356 & 0.964 &--3.91&--3.90&NGC 7654, double\\
\hline
\hline
\end{tabular}
  \\
  1 -- from the Hipparcos parallax (Paper1);
  2 -- from the cluster distance (Paper1);
  3 -- from the spectroscopic calibration of Paper1;
  4 -- \cite{ma06}
  \end{table*}

  \section{New luminosity calibrations}\label{cals}

  Figure 2 clearly shows that there is a correlation between the strength
  of the O{\sc i} triplet
  and the absolute magnitude. In the more luminous stars the triplet is
  significantly stronger.
  Nevertheless, there is a real scatter in this relation.
  To understand the source of the scatter, we plot the triplet
  strength versus
  the effective temperature in Fig. 3. Two different symbols (and two
  lines approximating their locii)
  represent two groups of supergiants according to the absolute
  magnitude: --3.5$<$\Mv$<$--2.5
  and        --5.0$<$\Mv$<$--4.0.
  It can be seen, that besides luminosity, the triplet also shows a strong dependence
  on ${T_{\rm eff}}$, particularly in the range 4500--6000 K,
  where it rapidly weakens towards the cooler side.
  However, at higher temperatures, the scatter is observed again,
  which implies dependence on even more parameters.
  For example, \cite{fa88} demonstrated the sensitivity
  of the triplet to the micro-turbulence.

  Using accurate photospheric parameters from Table 1
  for the 60 calibration stars with the known \Mv,  we derive the analytical
  expressions  (1) and (2) that relate the absolute magnitude \Mv\ to the triplet
  strength (EW71, EW74), \Teff, \Vt, \logg, and the iron abundance [Fe/H].

 \begin{eqnarray}
  M_{V}(1)&=&-2.73+17.14 \mu EW71 -0.425 \mu V_{t}  \nonumber\\
       &&+1.36 \mu \log g -24.41 EW71 +3.44 t [Fe/H] \nonumber\\
       &&+89.54 t EW71,
 \end{eqnarray}

\begin{eqnarray}
  M_{V}(2)&=&-2.77+5.79 \mu EW74 -0.53 \mu V_{t}    \nonumber\\
        &&+1.40 \mu \log g -7.53 EW74 + 4.29 t [Fe/H]  \nonumber\\
        &&+25.69 t EW74,
\end{eqnarray}

  where

 \begin{equation}
  \mu = e^{-(T_{eff} - 5340)^2 /(2 \times 1090^2)}
 \end{equation}

    and

 \begin{equation}
  t =  log(T_{eff})-3.7.
 \end{equation}

  Each expression consists of a linear combination of the
  products of the triplet EW and three photospheric parameters
  (\Vt, \logg, [Fe/H]), with two functions of \Teff ($\mu$ and $t$).
  This form is the result of the trial and error analysis, carried out with
  the goal to obtain precise,
  and at the same time clear and relatively compact expressions.
  The numerical coefficients have been obtained by using the generalized
  stepwise technique, and only significant terms have been retained in the final expressions.

  As has been mentioned earlier, the luminosity dependence of the triplet is
  the result of the non-LTE effects that depend on the physical
  parameters of the photosphere. Metallicity is one of such parameters,
  and for our stars without chemical anomalies, it
  can be represented simply by the iron abundance [Fe/H].
  Inclusion of the term with the oxygen abundance itself
  did not improve the precision of the calibrations.
  One reason is that with the EW exceeding 200 m\AA\,,
  the triplet falls past the linear regime on the curve of growth,
  being only weakly dependent on the abundance.
  The second reason is the lack of the strong deviation of [O/Fe]
  from the solar value in supergiants, of order $\pm$0.3 dex,
  which is comparable with the precision of the oxygen determination.
  Hence, it makes little difference (less than $\pm$0.2 mag in \Mv) whether to include the [Fe/H] or
  the [O/H] term in the equations.

  Using expressions (1) and (2), we obtained luminosities for
  all 74 F--G supergiants (\Mv (1) and \Mv (2)).
  The results are summarized in Table 1. Each entry includes:
  the name of the star,
  our photospheric model parameters, the strengths of the OI 7771.9 \AA\ line
  and of the 7774 \AA\ blend comprising all three components,
  the previously published and the new oxygen-based estimates of \Mv.
  As can be seen, the values obtained from the two oxygen indices are
  very close.
  For the stars with significantly broadened lines, the luminosity
  could only be
  determined from the second index, that relies on the EW of the whole feature.

  The absolute magnitudes recovered for the calibration stars are in excellent
  agreement with the original values from the literature, as shown in Fig. 4.
  The standard
  deviation is 0.26 mag with the correlation coefficient of 0.98.
  This is a significant improvement over the RMS$=$0.9 mag of the
  simple oxygen EW--\Mv\ relationship that ignores dependence on other parameters (Fig. 1).
  The accuracy of 0.26 mag achieved here, is primarily limited
  by the accuracy of the external methods of the luminosity determination,
  which will certainly be improved in the future.
  Based on the corresponding values in our sample,
  the applicable ranges of our calibrations are the following:
  \Teff$=$8000 to 4500 K (F0 to K0), \Mv$=$ --1 to --10 mag
  (luminosity classes I, II),
   and \FeH $=$--0.5 to $+$0.5.

  \begin{figure}
  \includegraphics[width=9.0cm]{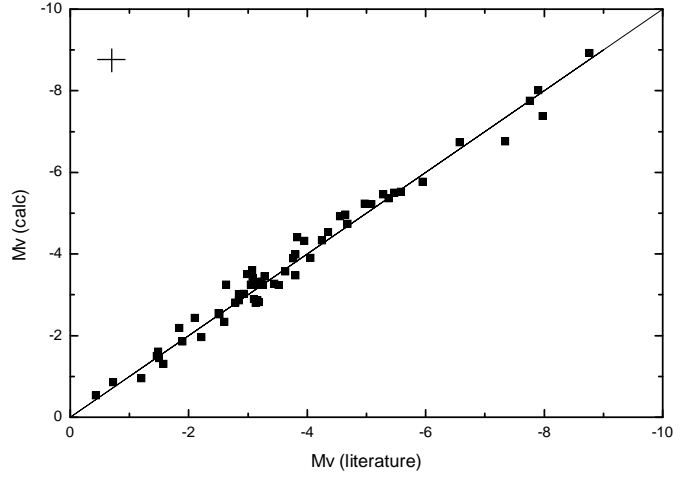}
  \caption{The comparison of our final calculated \Mv\ (an average value from the two oxygen indices)
                  with the estimates from the literature.
                  The line with a slope of 1 designates a perfect match.}
  \label{kovt4}
  \end{figure}

  \section{Summary}

  The location of a star (or a family of stars) in the HR diagram
  is a fundamental property required for the understanding of the
  structure and evolution of stars,
  since it enables a comparison with the evolutionary tracks and
  isochrones.
  Knowing the chemical composition, in addition, allows to obtain
  a deeper insight into the evolutionary processes that the star
  have undergone to achieve the present stage
  (e.g. \cite{lyu11}).
  The nature of the F--G supergiants studied here dictates that they are
  situated on average at much larger distances than dwarfs, and therefore,
  lack the parallax information for the direct estimate of the luminosity.
  In this study we present an alternative method for the luminosity
  determination  -- the accurate quantitative method based on the
  high-resolution spectra.

  We have refined the long-known method
  of the luminosity determination from the OI 7771--4 \AA\, triplet
  by incorporating the dependence on four other photospheric parameters:
  \Teff, \logg, \Vt, and [Fe/H] (or [O/H]).
  We derived two expressions that predict the absolute
  magnitudes of the F--G supergiants with precision of $\pm$0.26 mag.
  We note that these calibrations are only suitable for supergiants
  without any anomalies,
  such as binarity, chemical peculiarities, or the presence of the circumstellar
  matter (a possible indication of a low-mass star in the post-AGB stage).

  In the future, we plan to extend these
  calibrations on the classical Cepheids,
  which are the pulsating type of supergiants.
  This may help to disentangle the pulsation modes for the certain
  classes of Cepheids, particularly, for the short period ones like Polaris.
  At the same time, we plan to continue enlarging the number
  of supergiants with reliably and independently
  determined luminosities, which will enable to increase both the
  precision of the present calibrations and their application range.

 \section*{Acknowledgements}

 The spectra were collected with the ESO Telescopes at the
 Paranal Observatory under programme ID266.D-5655,
 and the Mercator Telescope, operated on the island
 of La Palma by the Flemish Community, at the Spanish
 Observatorio del Roque de los Muchachos of the Instituto de
 Astrofisica de Canarias.
 Much of the information about the supergiants was gathered
 with the help of SIMBAD.
 We thank the anonymous referee for the useful suggestions.

 \end{document}